\documentclass[%
 reprint,
 amsmath,amssymb,
 aps,
pre,
]{revtex4-1}

\usepackage{mathtools}
\usepackage{graphicx}% Include figure files
\usepackage{dcolumn}% Align table columns on decimal point
\usepackage{bm}% bold math
\usepackage{blindtext}
\usepackage{color}
\usepackage{footmisc}
\usepackage[caption=false]{subfig}
\usepackage[normalem]{ulem}

\newcommand{\vp}{\varphi}
\newcommand{\w}{\omega}
\newcommand{\W}{\Omega}
\newcommand{\e}{\varepsilon}
\newcommand{\te}{\widetilde{\varepsilon}}
\newcommand{\g}{\gamma}
\newcommand{\dd}{\mathrm{d}}

\begin{document}

%\preprint{APS/123-QED}

\title{Two Mechanisms of Remote Synchronization in a Chain of Stuart-Landau Oscillators}

\author{Mohit Kumar}
\email{mohitkumar.2k1@gmail.com}
\affiliation{Department of Mechanical Engineering, Indian Institute of Technology Madras, Chennai 600036, India}
\author{Michael Rosenblum}%
\email{mros@uni-potsdam.de}
\affiliation{Department of Physics and Astronomy, University of Potsdam, 
Karl-Liebknecht-Str. 24/25, D-14476 Potsdam-Golm, Germany
}

\date{\today}% It is always \today, today,

\begin{abstract}
Remote synchronization implies that oscillators interacting not directly but via an additional unit (hub) adjust their frequencies and exhibit frequency locking while the hub remains asynchronous. In this paper, we analyze the mechanisms of remote synchrony in a small network of three coupled Stuart-Landau oscillators using recent results on high-order phase reduction. We analytically demonstrate the role of two factors promoting remote synchrony. These factors are the non-isochronicity of oscillators and the coupling terms appearing in the second-order phase approximation. We show a good correspondence between our theory and numerical results for small and moderate coupling strengths.
\end{abstract}

\maketitle

\section{\label{sec:level1}Introduction}

Remote synchrony (RS) is an interesting manifestation of the general and highly significant nonlinear phenomenon of synchronization~\cite{Kuramoto-84,*pikovsky2003synchronization,*strogatz2004sync,*Osipov-Kurths-Zhou-07}. RS implies adjusting rhythms of oscillators that do not interact directly but only through an asynchronous unit (hub).  
Exploration of this effect, initially described by Bergner et al.~\cite{bergner2012RS} and further studied numerically and experimentally in Refs.~\cite{minati2015remote,*karakaya2019fading}, 
is crucial, e.g., for understanding functional connectivity in brain networks~\cite{vuksanovic2014functional,vlasov2017hub}. 

Previous studies analyzed RS in star-like and complex networks of Stuart-Landau (SL) or phase oscillators~\cite{bergner2012RS,minati2015remote,*karakaya2019fading,gambuzza2013analysis,vlasov2017hub}. The results uncovered the role of amplitude dynamics~\cite{bergner2012RS,gambuzza2013analysis}: 
RS appeared in a  network of isochronous SL units but not in its first-order phase approximation, i.e., in the Kuramoto network. 
Furthermore, Vlasov and Bifone~\cite{vlasov2017hub} demonstrated that RS emerges in 
networks of phase oscillators with the Kuramoto-Sakaguchi interaction~\cite{sakaguchi1986soluble}, but not in the case of zero phase shift in the sine-coupling term. Since the Kuramoto-Sakaguchi model is the first-order approximation of coupled non-isochronous SL oscillators, this result indicates the role of non-isochronicity in promoting RS.  
However, the understanding of mechanisms leading to RS is yet incomplete. This paper uses a simple motif of three coupled SL oscillators to analyze the transition to RS. In contradistinction to \cite{vlasov2017hub}, we consider non-identical peripheral oscillators.
Using recent results on high-order phase reduction~\cite{gengelphasereduction}, we explain the contribution of both the non-isochronicity and amplitude dynamics and quantitatively describe the transition to RS. We demonstrate the importance of high-order phase approximation in the explanation of RS. 

The paper is organized as follows. In Section~\ref{sec:model}, we introduce the 
model and its second-order phase approximation. Next, we demonstrate the transition to RS in this model. In Section~\ref{sec:RS_theoretical}, we derive the condition for this transition and 
in Section~\ref{sec:ref}, we present our results.
Section~\ref{sec:concl} concludes and discusses our findings.

\section{Remote synchrony in coupled Stuart-Landau oscillators}
\label{sec:model}
Consider three SL oscillators coupled in a chain as $1\xleftrightarrow{}2\xleftrightarrow{}3$. Thus, peripheral units 1 and 3 are not interacting directly but only through the central oscillator. 
Let the (generally different) natural frequencies of the oscillators be $\w_{1,2,3}$. Correspondingly, we denote the  frequencies of interacting units (observed frequencies) as $\W_{1,2,3}$. Following Bergner et al.~\cite{bergner2012RS}, we say that the network reaches a state of RS if, with an increase of coupling stength, $\W_1$ becomes equal to $\W_3$ while $\W_1\ne\W_2$. If all frequencies coincide, $\W_1=\W_2=\W_3$, then we speak about complete synchrony (CS).  
We emphasize that Refs.~\cite{qin2020mediated,*qin2018stability,*nicosia2013remote} use the term RS in a different context.

In the rest of this Section, we first specify our model and present its second-order phase approximation. Next, we numerically demonstrate transitions from asynchrony to RS and CS in the full model and its phase-reduced versions.

\subsection{Model and its phase approximation}

The governing equations of the model are:
\begin{equation}\label{eqn:SLequations}
    \dot{A}_{n}=\left[1+\mathrm{i} (\w_{n}+\alpha)\right]  A_{n}-(1+\mathrm{i}\alpha)\left|A_{n}\right|^{2} A_{n}+\e I_n \;,
\end{equation}
where $A_n \in \mathbb{C}$, $n=1,2,3$, $\w_n$ is the natural frequency of the $n$-th oscillator, 
and $\alpha$ is the non-isochronicity parameter, common for all units.   
The parameter $\e$ and the terms $I_1=A_2$, $I_2=A_1+A_3$, $I_3=A_2$ describe the strength and structure of the coupling, respectively.

It is well-known that for sufficiently weak coupling, the dynamics of interacting limit-cycle oscillators reduce to that of phases.  
For the coupled SL oscillators, the first-order phase approximation in $\e$ can be performed analytically because the phase of this system can be readily obtained from the state variable $A$; the reduction yields the celebrated Kuramoto-Sakaguchi phase equations \cite{sakaguchi1986soluble}.  
However, phase reduction beyond the first-order approximation remains challenging and is a subject of ongoing research.  Here, we use the results of Gengel et al.~\cite{gengelphasereduction}, who 
provided expressions for the second-order reduction of coupled SL oscillators~\footnote{Notice that our Eq.~(\ref{eqn:SLequations}) represents a particular case of a more general setup studied in~\cite{gengelphasereduction}}.
Let the phase of the $n$-th oscillator be $\vp_n$. The second-order phase approximation of the system (\ref{eqn:SLequations}) reads: 
\begin{widetext}
\begin{equation}
\begin{aligned}\label{eqn:second_order_phase_reduction}
\dot{\vp}_{1}=& \w_{1}+\e \left[\sin \left(\vp_{2}-\vp_{1}\right)-\alpha \cos \left(\vp_{2}-\vp_{1}\right)\right]\\
&+\e^{2}\left[D_{32} \cos \left(2 \vp_{2}-\vp_{1}-\vp_{3}\right)+C_{32} \sin \left(2 \vp_{2}-\vp_{1}-\vp_{3}\right)-D_{32} \cos \left(\vp_{3}-\vp_{1}\right)+C_{32} \sin \left(\vp_{3}-\vp_{1}\right)\right] +\mathcal{O}(\e^3)\;, \\
\dot{\vp}_{2}=& \w_{2}+\e \left[\sin \left(\vp_{1}-\vp_{2}\right)-\alpha \cos \left(\vp_{1}-\vp_{2}\right)+\sin \left(\vp_{3}-\vp_{2}\right)-\alpha \cos \left(\vp_{3}-\vp_{2}\right)\right] \\
&+\e^{2}\left[(D_{12}+D_{32}) \cos \left(2 \vp_{2}-\vp_{1}-\vp_{3}\right)+(C_{12}+C_{32}) \sin \left(2 \vp_{2}-\vp_{1}-\vp_{3}\right)\right.\\
&-\left.(D_{12}+D_{32}) \cos \left(\vp_{1}-\vp_{3}\right)+(C_{12}-C_{32})\sin(\vp_1 - \vp_3)\right] +\mathcal{O}(\e^3)\;,\\
\dot{\vp}_{3}=& \w_{3}+\e \left[\sin \left(\vp_{2}-\vp_{3}\right)-\alpha \cos \left(\vp_{2}-\vp_{3}\right)\right] \\
&+\e^{2}\left[D_{12} \cos \left(2 \vp_{2}-\vp_{3}-\vp_{1}\right)+C_{12} \sin \left(2 \vp_{2}-\vp_{3}-\vp_{1}\right) -D_{12} \cos \left(\vp_{1}-\vp_{3}\right)+C_{12} \sin \left(\vp_{1}-\vp_{3}\right)\right]+\mathcal{O}(\e^3)\;,
\end{aligned}
\end{equation}
\end{widetext}
where 
\begin{equation}\label{eqn:second_order_phase_reduction_constant_C}
    C_{ij}=\frac{1+\alpha^2}{4+(\w_i-\w_j)^2}
\end{equation}
and
\begin{equation}\label{eqn:second_order_phase_reduction_constant_D}
    D_{ij}=\frac{1+\alpha^2}{2}\left(\frac{\w_i-\w_j}{4+(\w_i-\w_j)^2} \right)\;.
\end{equation}

Keeping in Eq.~(\ref{eqn:second_order_phase_reduction}) only the first-order terms $\sim\e$, one obtains the Kuramoto-Sakaguchi model:
\begin{equation}
\begin{aligned}\label{eqn:first_order_phase_reduction}
\dot{\vp}_{1}=& \w_{1}+\e \left[\sin \left(\vp_{2}-\vp_{1}\right)-\alpha \cos \left(\vp_{2}-\vp_{1}\right)\right]\;, \\
\dot{\vp}_{2}=& \w_{2}+\e \left[\sin \left(\vp_{1}-\vp_{2}\right)-\alpha \cos \left(\vp_{1}-\vp_{2}\right)\right.\\
&+\left.\sin \left(\vp_{3}-\vp_{2}\right)-\alpha \cos \left(\vp_{3}-\vp_{2}\right)\right] \;, \\
\dot{\vp}_{3}=& \w_{3}+\e \left[\sin \left(\vp_{2}-\vp_{3}\right)-\alpha \cos \left(\vp_{2}-\vp_{3}\right)\right] \;.
\end{aligned}
\end{equation}
For isochronous oscillators, $\alpha=0$, the model simplifies to the Kuramoto network.

\subsection{Remote synchrony in the full and reduced models}
\label{sec:RS_numerical}

\begin{figure}
    \includegraphics[width=\columnwidth]{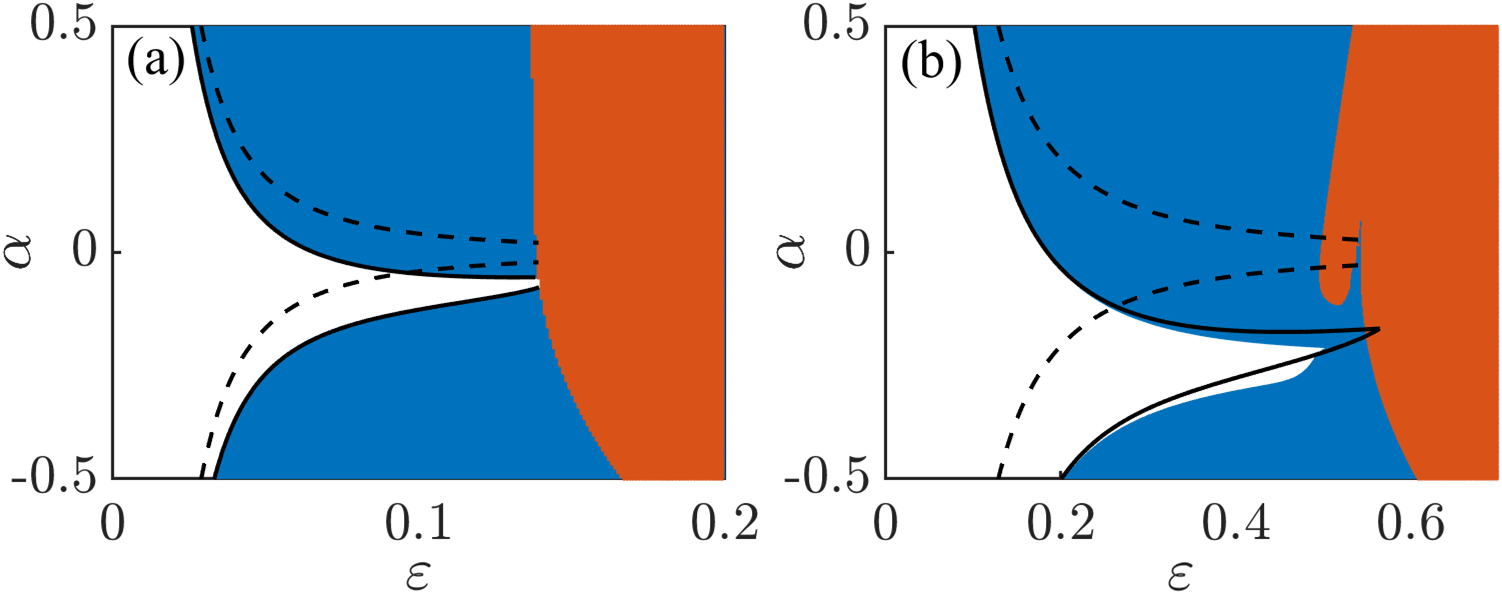}
    \caption{\label{fig:compare_approximations_figures}
    (Color online) Numerically computed bifurcation diagrams illustrating the dependence of the system's observed state on the coupling strength, $\e$, and non-isochronicity parameter, $\alpha$. The oscillators' natural frequencies are (a) $\w_1=1$, $\w_2=\sqrt{2}$, $\w_3=1.002$ and (b) $\w_1=1$, $\w_2=\sqrt{7}$, $\w_3=1.01$.
    The white, blue (dark gray), and red (light gray) regions correspond to asynchrony, RS, and CS, respectively, upon numerical simulation of  Eq.~\eqref{eqn:SLequations}. The solid black line depicts the RS transition border as computed using the second-order phase approximation, see Eqs.~\eqref{eqn:second_order_phase_reduction}. The dashed black line shows the RS transition obtained for the first-order phase approximation, see
    Eqs.~\eqref{eqn:first_order_phase_reduction}. The diagrams demonstrate the crucial role of the non-isochronicity parameter $\alpha$. Furthermore, the diagrams clearly show the advantage of the second-order approximation.
    }
\end{figure}

This section compares and contrasts the regions of RS obtained using the SL system~\eqref{eqn:SLequations} and the phase approximations, see Eqs.~(\ref{eqn:first_order_phase_reduction},\ref{eqn:second_order_phase_reduction}). To this end, we fix the natural frequencies of all three oscillators~\footnote{In the following, we always consider frequencies of the peripheral oscillators to be close while the frequency of the hub is essentially different.}, numerically simulate the governing equations, and detect regions of asynchrony, CS and RS upon varying the coupling strength and the non-isochronicity parameter. (The description of the numerical procedures are deferred to Section \ref{sec:ref}.) This results in two-parameter bifurcation diagrams on the $\e$-$\alpha$ plane shown in Fig.~\ref{fig:compare_approximations_figures}.

Figure \ref{fig:compare_approximations_figures} provides us with two insights. Firstly, we note that the first-order approximation does not accurately reproduce the transition to RS. This approximation's failure results from not accounting for the amplitude modulation in the coupled SL oscillators. On the other hand, the second-order approximation fares well and is accurate for small and moderate coupling strengths. 
% This discussion highlights the role of amplitude dynamics in inducing RS, as pointed out in Refs. \cite{bergner2012RS,gambuzza2013analysis}. 
Secondly, the non-isochronicity parameter essentially affects the transition to RS.  Generally, RS in the SL system \eqref{eqn:SLequations} appears for both the isochronous ($\alpha=0$) and the non-isochronous ($\alpha\ne 0$) cases. However, this feature is captured only by the second-order approximation; the first approximation does not exhibit RS for $\alpha=0$, in agreement with the results by Vlasov and Bifone \cite{vlasov2017hub}.

\section{Theoretical analysis of the phase dynamics}
\label{sec:RS_theoretical}
We use the phase equations~\eqref{eqn:second_order_phase_reduction}
to investigate the transition to RS.  It is straightforward to reduce Eqs.~\eqref{eqn:second_order_phase_reduction} 
to a two-dimensional system for the phase differences: 
\begin{equation}\label{eqn:define_phase_differences}
    \g_{13}=\vp_1 - \vp_3, \quad \g_{12}=\vp_1 - \vp_2 \;.
\end{equation}
The resulting equations represent the dynamics on a two-torus and can be studied using standard phase plane analysis techniques. In terms of the phase differences, the asynchronous state corresponds to an unbounded growth (or decline) of $\g_{13}$ and $\g_{12}$. Upon increasing the coupling strength, one observes RS, wherein $\g_{13}$ is bounded while $\g_{12}$ is unbounded. For transparency and brevity, we present our theory by analyzing the first-order phase equations. Then we provide the results of the same approach applied to the second-order model. 

\begin{figure}[t]
\includegraphics[width=\columnwidth]{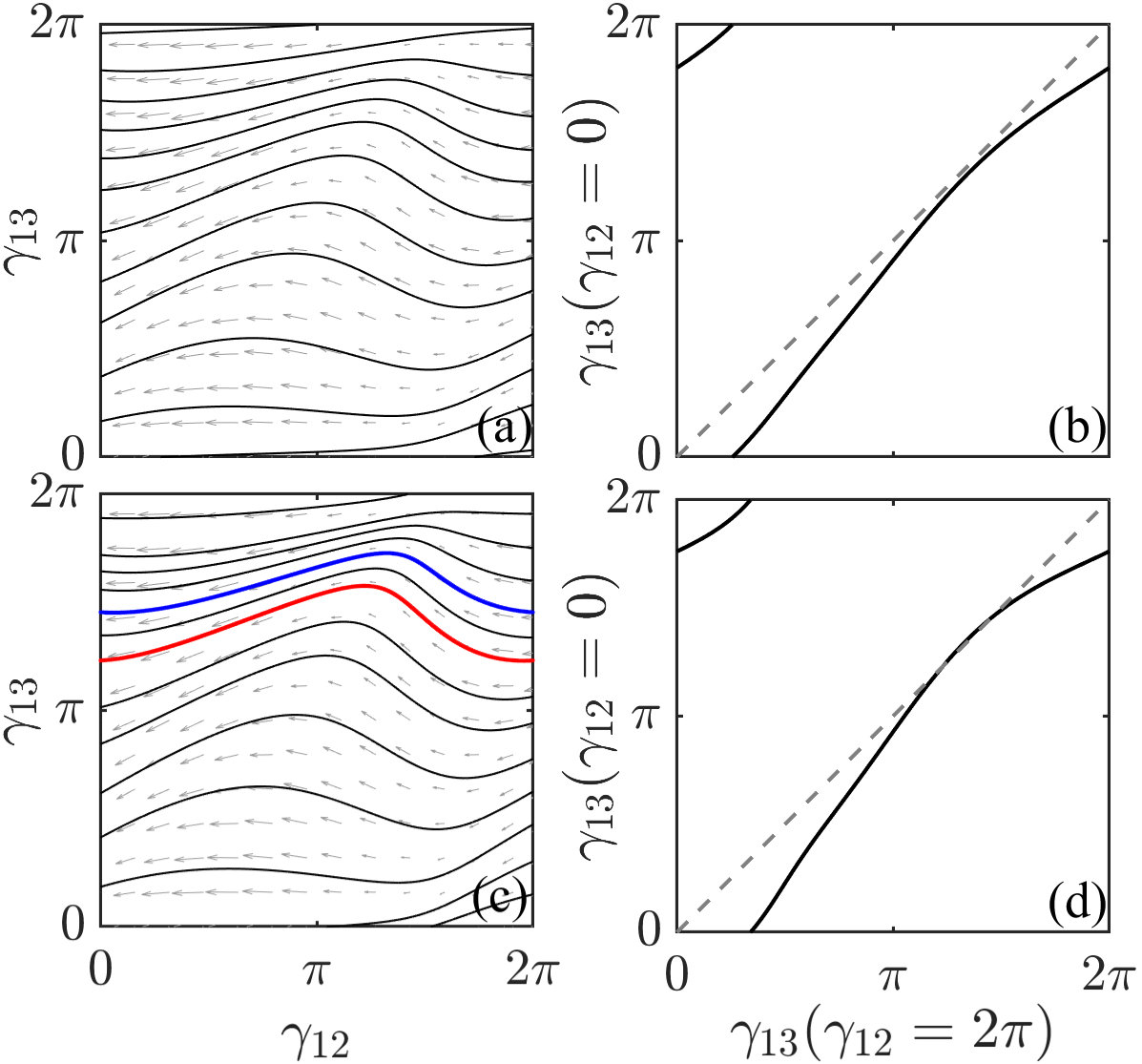}
\caption{% \label{fig:RS_transition_figures}
(Color online) Illustration of the RS transition using the phase approximation equations upon increasing the coupling strength.
(a) and (c): Phase plane representations of the phase difference dynamics at low and moderate coupling strengths, respectively. The bold blue (dark gray) and red (light gray) trajectories in panel (c) depict the stable and unstable limit cycles.
(b) and (d): Poincar\'e maps corresponding to the flows shown in panels (a) and (c), respectively, constructed using the Poincar\'e section $\gamma_{12}=2\pi$. 
The birth of a stable limit cycle (fixed point) in the phase plane plot  (Poincar\'e map) corresponds to the onset of RS.
(The plots shown here are computed using the first-order approximation [Eq.~\eqref{eqn:first_order_phase_reduction}] but they are qualitatively identical for the second-order phase reduction case as well.)}
\label{fig:model}
\end{figure}

\subsection{Poincar\'e map}

The transition to RS corresponds to the appearance of a stable limit cycle (LC) on the torus. Figure 2a depicts a typical situation for the asynchronous regime at low coupling strengths. There are no attractors on the phase plane, the motion is quasiperiodic, and the phase differences $\g_{13}$ and $\g_{12}$ are unbounded. Figure 2c exemplifies the RS state once the coupling strength increases. A stable and an unstable limit cycle are born via a saddle-node bifurcation of LCs. Notice that on the LC, $\g_{12}$ is unbounded while $\g_{13}$ is bounded, which indicates the emergence of RS. Notice also that we consider $\w_1<\w_2$ for definiteness for the remainder of this article. Hence, the flow is from right to left. We have verified that our conclusions hold equally well for the other case.

For the following derivation, it is instructive to construct a Poincar\'e map, choosing the line $\g_{12}=2\pi$  as the Poincar\'e section. A trajectory that begins on this section intersects it next at $\g_{12}=0$, since the flow on the torus is leftwards. Thus, we have  $\g_{13}(0)=P(\g_{13}(2\pi))$, where $P(\cdot)$ denotes the Poincar\'e map. The Poincar\'e map corresponding to Figs.~2a and 2c are shown in Figs.~2b and 2d, respectively. Evidently, RS in the system equates to a stable fixed point of the Poincar\'e map. We exploit this observation to derive the condition for RS analytically.

\subsection{First-order phase dynamics}
Starting with Eqs.~\eqref{eqn:first_order_phase_reduction}, using Eq.~\eqref{eqn:define_phase_differences}, 
and introducing the new time $\tau=(\w_2-\w_1) t$,
we obtain a two-dimensional system for phase differences:
\begin{equation}
    \begin{aligned}\label{eqn:phase_differences_first_order_phase_reduction}
    \g'_{13}&=\nu+\te\left[ -\sin\g_{12}-\alpha\cos\g_{12} - \sin(\g_{13}- \g_{12})\right. \\
    &\quad +\left.\alpha \cos (\g_{13} -\g_{12}) \right]\;,\\
    \g'_{12}&=-1+\te\left[ -2\sin\g_{12} - \sin(\g_{12} -\g_{13})\right.\\
    & \quad + \left.\alpha\cos(\g_{12} -\g_{13}) \right]\;,
    \end{aligned}
\end{equation}
where
\begin{equation}
\begin{gathered}\label{eqn:my_params}
    \nu=\frac{\w_1 - \w_3}{\w_2-\w_1}\;, \quad \te=\frac{\e}{\w_2-\w_1}\;,
\end{gathered}
\end{equation}
and $(\cdot)'$ denotes differentiation with respect to $\tau$.

To derive the Poincar\'e map $\g_{13}(0)=P(\g_{13}(2\pi))$, we divide the preceding equations to obtain:
\begin{widetext}
\begin{equation}\label{eqn:first_order_gamma13/gamma12}
\begin{aligned}
    \frac{\dd \g_{13}}{\dd\g_{12}} &= \frac{\nu+\te\left[ -\sin\g_{12}-\alpha\cos\g_{12} - \sin(\g_{13}- \g_{12}) +\alpha \cos (\g_{13} -\g_{12}) \right]}{-1+\te\left[ -2\sin\g_{12} - \sin(\g_{12} -\g_{13}) + \alpha\cos(\g_{12} -\g_{13}) \right]}\;.
\end{aligned}
\end{equation}
\end{widetext}
We solve Eq.~\eqref{eqn:first_order_gamma13/gamma12} with the initial condition $\g_{13}(2\pi)$  using a perturbation approach, for which we assume the following:
\begin{equation}\label{eqn:assumptions_sys_params}
    |\w_1 - \w_2| \sim \mathcal{O}(1),\quad 0<|\w_1 - \w_3| \ll 1, \quad \e \ll 1\;.
\end{equation}
Notice that the first pair of assumptions formally encapsulates our previous qualitative description: the peripheral oscillators are near-identical, whereas the hub oscillator is markedly different. Equivalently, in terms of the parameters present in Eq.~\eqref{eqn:phase_differences_first_order_phase_reduction}, the assumptions result in 
$\te\ll 1$ and $\nu \ll 1$.

The solution presented in Appendix~\ref{app:1st_order} provides the condition for the existence of the Poincar\'e map's fixed point:
\begin{equation}\label{eqn:first_order_RS_condition}
    \left| \frac{\e^2 \alpha}{(\w_1-\w_3)(\w_1-\w_2)} \right| \geq \frac{1}{2}\;.
\end{equation}
This inequality yields the necessary condition for RS in the first-order phase reduction Eqs.~\eqref{eqn:first_order_phase_reduction}. Its validity depends on the smallness of $\e$. It indicates that upon increasing the coupling strength, RS appears due to non-isochronicity. Hence, RS is impossible in a chain of three non-identical Kuramoto equations. This result agrees with the observation reported in 
Ref.~\cite{bergner2012RS} and theoretical analysis in Ref.~\cite{vlasov2017hub}. 

\subsection{Second-order phase dynamics}
Now, we use the same technique to construct the Poincar\'e map from the second-order phase dynamics equations. For this goal, we re-write Eqs.~\eqref{eqn:second_order_phase_reduction} in terms of phase differences and then obtain an equation for $\frac{\dd\gamma_{13}}{\dd\gamma_{12}}$ that is similar to Eq.~\eqref{eqn:first_order_gamma13/gamma12} but contains 
additional terms proportional to $\te^2$. Solving this equation by the perturbation technique (see Appendix~\ref{app:2nd_order} for details), we arrive at the following condition for RS:
\begin{equation}
\label{eqn:second_order_RS_condition}
    \left| \frac{{\e}^2 [\alpha-(\w_1 - \w_2)C_{12}]}{(\w_1-\w_3)(\w_1-\w_2)} \right| \geq \frac{1}{2}\;.
\end{equation}
This condition differs from the inequality \eqref{eqn:first_order_RS_condition}, derived in the first approximation, by the term $(\w_1 - \w_2)C_{12}$ alone. 
(Notice that $C_{12}\approx C_{32}$.) This term is proportional to the amplitude of the synchronizing 
terms $\sin(\vp_3-\vp_1)$, $\sin(\vp_1-\vp_3)$ in~Eqs.~\eqref{eqn:second_order_phase_reduction}.
These terms indicate the presence of an ``invisible'' coupling between oscillators 1 and 3. This coupling exists despite the absence of a physical link between the first and third units; the first-order phase reduction does not reveal it. Thus, RS is promoted by non-isochronicity and by indirect coupling through the hub.

\section{Results}
\label{sec:ref}

To validate our derivations, we compare the bifurcation diagram on the $\e$-$\alpha$ plane obtained using the various approximations against those obtained for the exact SL equations. Before discussing the plots, we briefly recall the approximations made and clarify the terminology used to distinguish between them. The results from the numerical computations using the SL system~\eqref{eqn:SLequations} will be referred to as ``exact''. If the numerical calculation used the first-order [Eq.~\eqref{eqn:first_order_phase_reduction}] or the second-order 
[Eq.~\eqref{eqn:second_order_phase_reduction}] phase reduction, the corresponding result will be termed as ``NPR1'' or ``NPR2'', respectively~\footnote{We sweep the parameter space to determine the state of the system using efficient techniques. We find RS in the SL equations~\eqref{eqn:SLequations} by looking for a limit cycle solution where $\g_{12}$ is unbounded while $\g_{13}$ is bounded using a shooting method. To detect RS using the phase reduction equations exactly, we numerically construct the Poincar\'e map described in Sec. \ref{sec:RS_theoretical} by simulating Eq.~\eqref{eqn:first_order_gamma13/gamma12} (or its second-order counterpart) and check the presence of a fixed point. Regions of CS were computed using direct numerical simulations; we mark the points in the parameter space that resulted in $\Omega_1=\Omega_2=\Omega_3$ up to a tolerance of $10^{-4}$.}. Finally, the theoretical results obtained for the first-order [Eq. \eqref{eqn:first_order_RS_condition}] and second-order [Eq. \eqref{eqn:second_order_RS_condition}] phase reduction are coined as ``TPR1'' and ``TPR2'', respectively~\footnote{The borderline of the RS transition is given by the condition when inequalities (\ref{eqn:first_order_RS_condition},\ref{eqn:second_order_RS_condition}) turn to equalities.}.

As a first step, we compared the NPR1 and TPR1 borderlines of the RS transitions. We found that TPR1 very well reproduces the numerical results shown by dashed lines in Fig.~\ref{fig:compare_approximations_figures}.
This result confirms the capability of the perturbation approach to capture RS in the Kuramoto-Sakaguchi model \eqref{eqn:first_order_phase_reduction}. 
%However, as shown in Sec. \ref{sec:RS_numerical}, this approximation fails against the SL system. 

\begin{figure}
    \includegraphics[width=\columnwidth]{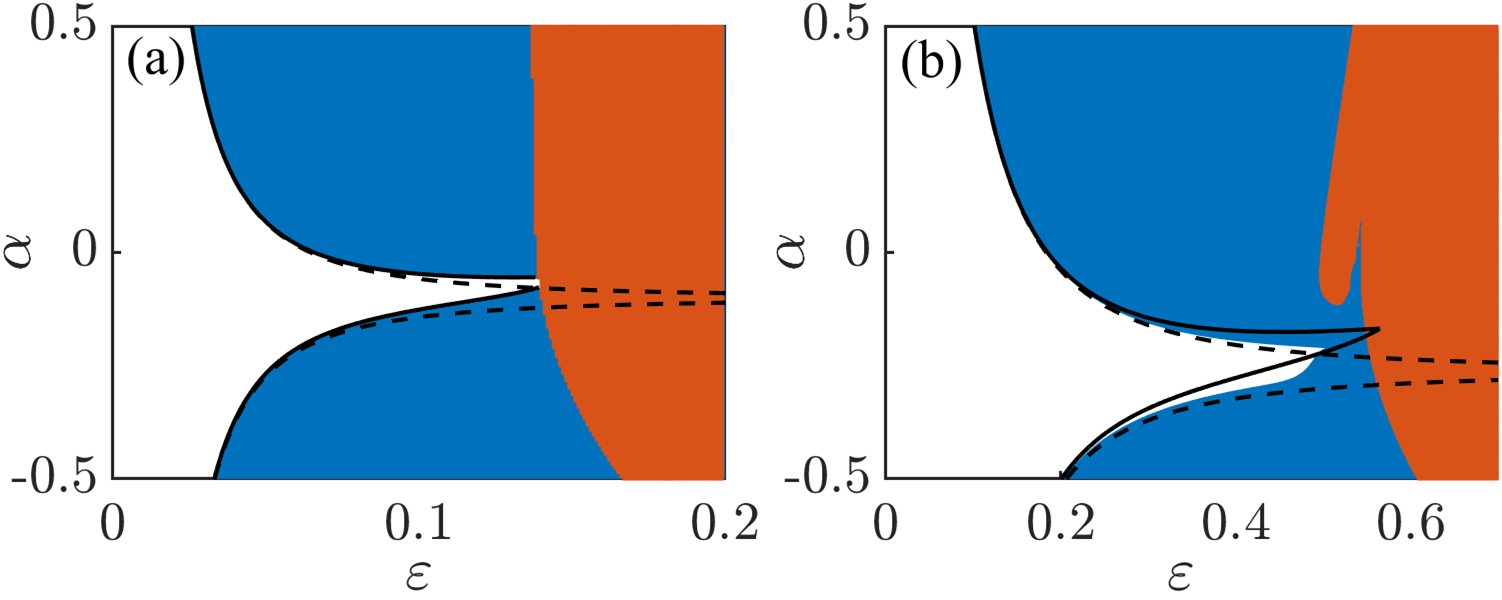}
    \caption{\label{fig:secondorder_figures}
    (Color online) Comparison of theoretical and numerical results. 
    Two-parameter bifurcation diagrams on the $\e$-$\alpha$ plane (coupling strength versus non-isochronicity) depicting the system's state. Exact domains of RS, CS, and asynchrony are shown in blue (dark gray), red (light gray), and white, respectively. The solid black line shows the RS borderline obtained numerically using the second-order phase reduction (NPR2). The dashed black line is the corresponding theoretical solution (TPR2). The oscillators' natural frequencies are (a) $\w_1=1$, $\w_2=\sqrt{2}$, $\w_3=1.002$ and (b) $\w_1=1$, $\w_2=\sqrt{7}$, $\w_3=1.01$.}
\end{figure}

Figure~\ref{fig:secondorder_figures} presents our main result. Here, we compare the NPR2 and TPR2 borderlines of the RS transition against the exact ones. When the frequency detuning $|\w_1-\w_3|$ is very small, as in Fig.~3a, all borders are practically identical for low coupling strengths. As the coupling strength $\e$ increases, the normalized coupling $\te$ (see Eq.~\eqref{eqn:my_params}) is no longer small, which causes the observed deviation between the TPR2 and NPR2 borders. Note that the NPR2 border accurately reproduces the exact RS transition throughout the considered range of coupling strengths. 
The bifurcation diagram for a second set of natural frequencies is presented in Fig.~3b. Again, for low coupling strengths, the agreement between the approximations and the exact solution is perfect. However, both NPR2 and TPR2 borders deviate from the exact border of the RS transition for higher values of coupling strength. This deviation occurs because $\e$ (and likewise $\te$) are no longer small quantities. We mention in passing that the dynamics for higher coupling strengths is often not trivial. For instance, the transition to CS in Fig.~3b near the finger-like structure around the point $(\e=0.5,\alpha=0)$ exhibits complex, possibly chaotic, dynamics, presumably due to the effects of strong coupling. Interestingly, near this point, there exists a window of RS straddled by regions of CS on either side.  

\section{Conclusions}
\label{sec:concl}
In summary, we analyzed the mechanisms of RS in a chain of three SL oscillators. We demonstrated that the RS transition is determined by the interplay of the non-isochronicity and the amplitude dynamics. The impact of the latter factor renders the standard first-order phase dynamics description of the RS phenomenon invalid. Our result emphasizes the importance of high-order phase reduction and highlights the crucial role amplitude dynamics may have in governing the behavior of networks of nonlinear oscillators. 

We believe that the effect of the amplitude dynamics neglected in the first-order phase approximation and revealed by the high-order one holds for general limit-cycle oscillators.
This belief is supported by the results of numerical network reconstruction from data~\cite{Kralemann-Pikovsky-Rosenblum-11}, which demonstrated the emergence of coupling between indirectly interacting units. It will be interesting to investigate how the unit's complexity may bring about qualitatively new changes to the RS transition~\cite{lacerda2019multistable,minati2015remote,*karakaya2019fading} and if they can be explained under the present framework.  

\begin{acknowledgments}
MK is grateful for the WISE scholarship by the DAAD (German Academic Exchange Service), which facilitated this work.
\end{acknowledgments}

\appendix

\section{Perturbative solution for the first-order phase approximation}
\label{app:1st_order}

Let us assume a power series expansion for $\g_{13}(\g_{12})$ in $\te$ as follows:
\begin{equation}\label{eqn:gamma13_series}
    \g_{13}(\g_{12})=\g_{13;0}(\g_{12})+\te \g_{13;1}(\g_{12}) + \te^2 \g_{13;2}(\g_{12}) + \mathcal{O}(\te^3)\;.
\end{equation}
The next step is to substitute this expansion in Eq.~\eqref{eqn:first_order_gamma13/gamma12} and gather the terms with matching powers of $\te$. However, it is unclear where the terms involving $\nu$ shall be grouped, as the relation between $\nu$ and $\te$ is unknown. This is not a problem since we may arbitrarily assume any order for $\nu$; its correct scaling near the RS transition is found as part of the derivation by the principle of dominant balance \cite{miller2006applied}. For illustration, we have grouped $\nu$ with the $\mathcal{O}(1)$ terms. (Alternatively, one may want to group it with $\mathcal{O}(\te^2)$ terms as that makes Eqs.~\eqref{eqn:collected_terms_first_order_perturbation} shorter.) Now, we collect the terms at each order as follows:
\begin{widetext}
\begin{equation}
\begin{aligned}\label{eqn:collected_terms_first_order_perturbation}
    \mathcal{O}(\te^0)&:  &\frac{\dd \g_{13;0}}{\dd\g_{12}}=&-\nu\;,\\
    \mathcal{O}(\te^1)&: &\frac{\dd \g_{13;1}}{\dd\g_{12}}=&-\alpha  (\nu +1) \cos \left(\g_{12}-\g
   _{13;0}\right)+\alpha  \cos \left(\g_{12}\right)+2 \nu  \sin
   \left(\g_{12}\right)+\nu  \sin \left(\g_{12}-\g
   _{13;0}\right)+\sin \left(\g_{12}\right)\\
   & & & -\sin \left(\g_2-\g _{13;0}\right)\;,\\
   \mathcal{O}(\te^2)&: &\frac{\dd \g_{13;2}}{\dd\g_{12}}=&\left(\alpha  \cos \left(\g_{12}-\g_{13;0}\right)-2 \sin
   \left(\g_{12}\right)-\sin \left(\g_{12}-\g_{13;0}\right)\right) \left(-\alpha  (\nu +1) \cos \left(\g_{12}-\g_{13;0}\right)+\alpha  \cos \left(\g_{12}\right)\right.\\
   & & & \left.+2\nu  \sin \left(\g_{12}\right) +\nu  \sin \left(\g_{12}-\g_{13;0}\right)+\sin \left(\g_{12}\right)-\sin\left(\g_{12}-\g_{13;0}\right)\right)\\
   & & &-\g_{13;1}
   \left(\alpha  (\nu +1) \sin \left(\g_{12}-\g_{13;0}\right)+(\nu -1) \cos \left(\g_{12}-\g_{13;0}\right)\right)\;.
\end{aligned}
\end{equation}
\end{widetext}
The initial conditions associated with the differential equation of each order are:
\begin{equation}\label{eqn:series_ICs}
    \g_{13;0}(2\pi)=\g_{13}(2\pi),\quad \g_{13;1}(2\pi)=0,\quad \g_{13;2}(2\pi)=0\;.
\end{equation}
Equations~\eqref{eqn:collected_terms_first_order_perturbation} along with the initial conditions in Eqs.~\eqref{eqn:series_ICs} are solved sequentially, providing the solutions for $\g_{13;0}$, $\g_{13;1}$ and $\g_{13;2}$. These terms are now substituted back into the series expansion Eq.~\eqref{eqn:gamma13_series}. By evaluating the resultant expression at $\g_{12}=0$, we arrive at a functional relation between $\g_{13}(2\pi)$ and $\g_{13}(0)$, which is the desired Poincar\'e map. The described procedure yields:  
\begin{equation}
\begin{aligned}
    \g_{13}(\g_{12})&=\g_{13;0}(\g_{12};\g_{13}(2\pi))+\te \g_{13;1}(\g_{12};\g_{13}(2\pi))\\
    &  + \te^2 \g_{13;2}(\g_{12};\g_{13}(2\pi))+\mathcal{O}(\te^3)\;,\\
    \g_{13}(0)&=\g_{13;0}(0;\g_{13}(2\pi))+\te \g_{13;1}(0;\g_{13}(2\pi))\\
    &  + \te^2 \g_{13;2}(0;\g_{13}(2\pi))+\mathcal{O}(\te^3)=P(\g_{13}(2\pi))\;,
\end{aligned}
\end{equation}
where the solution's dependence on the initial condition $\g_{13}(2\pi)$ has been explicitly pointed out using a semicolon notation.

With the expression for the Poincar\'e map derived, the final step involves solving for the map's fixed points. Evaluating the expression $P(\g_{13}(2\pi))=\g_{13}(2\pi)$ leads to:
\begin{equation}\label{eqn:first_order_eqn_condition}
    \nu - 2 \te^2 \alpha \sin (\g_{13}(2\pi)) + \mathcal{O}(\te\nu)=0 \;.
\end{equation}
(By the principle of dominant balance, Eq.~\eqref{eqn:first_order_eqn_condition} indicates that $\nu \sim \mathcal{O}(\te^2)$. Thus, we have found the correct scaling for $\nu$ in the neighbourhood of RS.)
Upon neglecting the higher-order terms, the preceding equation is tantamount to:
\begin{equation}
    \sin(\g_{13}(2\pi)) = \frac{\nu}{2 \te^2 \alpha}\;.
\end{equation}
For the above equation to have a solution, the absolute value of the right-hand side must be lesser than unity. This gives:
\begin{equation}
    \left| \frac{ \te^2 \alpha}{\nu} \right| \geq \frac{1}{2} \;.
\end{equation}
Finally, we revert back to our original parameters $\w_1$, $\w_2$, and $\e$ using Eq.~\eqref{eqn:my_params} to obtain:
\begin{equation}
%\label{eqn:first_order_RS_condition}
    \left| \frac{\e^2 \alpha}{(\w_1-\w_3)(\w_1-\w_2)} \right| \geq \frac{1}{2}\;.
\end{equation}

\section{Perturbative solution for the second-order phase approximation}
\label{app:2nd_order}

This Appendix derives the condition for RS using the second-order phase approximation. As done earlier, we exploit the assumptions formulated in Eq.~\eqref{eqn:assumptions_sys_params}. This allows us to simplify Eq.~\eqref{eqn:second_order_phase_reduction} as follows:
\begin{widetext}
\begin{equation}
\begin{aligned}\label{eqn:second_order_phase_reduction_simplified}
\dot{\vp}_{1}=& \w_{1}+\e \left[\sin \left(\vp_{2}-\vp_{1}\right)-\alpha \cos \left(\vp_{2}-\vp_{1}\right)\right] \\
&+\e^{2}\left[D_{12} \cos \left(2 \vp_{2}-\vp_{1}-\vp_{3}\right)+C_{12} \sin \left(2 \vp_{2}-\vp_{1}-\vp_{3}\right)-D_{12} \cos \left(\vp_{3}-\vp_{1}\right)+C_{12} \sin \left(\vp_{3}-\vp_{1}\right)\right]\;, \\
\dot{\vp}_{2}=& \w_{2}+\e \left[\sin \left(\vp_{1}-\vp_{2}\right)-\alpha \cos \left(\vp_{1}-\vp_{2}\right)+\sin \left(\vp_{3}-\vp_{2}\right)-\alpha \cos \left(\vp_{3}-\vp_{2}\right)\right] \\
&+\e^{2}\left[2D_{12} \cos \left(2 \vp_{2}-\vp_{1}-\vp_{3}\right)+2C_{12} \sin \left(2 \vp_{2}-\vp_{1}-\vp_{3}\right)-2D_{12} \cos \left(\vp_{1}-\vp_{3}\right)\right]\;, \\
\dot{\vp}_{3}=& \w_{3}+\e \left[\sin \left(\vp_{2}-\vp_{3}\right)-\alpha \cos \left(\vp_{2}-\vp_{3}\right)\right] \\
&+\e^{2}\left[D_{12} \cos \left(2 \vp_{2}-\vp_{3}-\vp_{1}\right)+C_{12} \sin \left(2 \vp_{2}-\vp_{3}-\vp_{1}\right) -D_{12} \cos \left(\vp_{1}-\vp_{3}\right)+C_{12} \sin \left(\vp_{1}-\vp_{3}\right)\right]\;,
\end{aligned}
\end{equation}
\end{widetext}
where $C_{ij}$ and $D_{ij}$ were defined in Eqs.~\eqref{eqn:second_order_phase_reduction_constant_C} and \eqref{eqn:second_order_phase_reduction_constant_D}. In particular, we have used $C_{32} \approx C_{12}$ and $D_{32} \approx D_{12}$ (up to the second order). Notice the presence of terms of the form $\sin(\vp_1 - \vp_3)$ in the first and last of Eqs.~\eqref{eqn:second_order_phase_reduction_simplified}, which explicitly indicate the ``invisible'' coupling between oscillators 1 and 3.

Hereafter, the procedure to derive the criteria for RS is identical to that of the first-order approximation and is not presented here for brevity. The expression obtained upon solving for the fixed points of the Poincar\'e map is: 
\begin{equation}\label{eqn:second_order_eqn_condition}
    \nu - 2 \te^2 (\alpha - (\w_1 - \w_2)C_{12})\sin (\g_{13}(2\pi)) + \mathcal{O}(\te\nu)=0\;,
\end{equation}
which has a solution for $\g_{13}(2\pi)$ if:
\begin{equation}
%\label{eqn:second_order_RS_condition}
    \left| \frac{{\e}^2 [\alpha-(\w_1 - \w_2)C_{12}]}{(\w_1-\w_3)(\w_1-\w_2)} \right| \geq \frac{1}{2}\;.
\end{equation}

%\bibliography{References}% Produces the bibliography via BibTeX.

\begin{thebibliography}{22}%
\makeatletter
\providecommand \@ifxundefined [1]{%
 \@ifx{#1\undefined}
}%
\providecommand \@ifnum [1]{%
 \ifnum #1\expandafter \@firstoftwo
 \else \expandafter \@secondoftwo
 \fi
}%
\providecommand \@ifx [1]{%
 \ifx #1\expandafter \@firstoftwo
 \else \expandafter \@secondoftwo
 \fi
}%
\providecommand \natexlab [1]{#1}%
\providecommand \enquote  [1]{``#1''}%
\providecommand \bibnamefont  [1]{#1}%
\providecommand \bibfnamefont [1]{#1}%
\providecommand \citenamefont [1]{#1}%
\providecommand \href@noop [0]{\@secondoftwo}%
\providecommand \href [0]{\begingroup \@sanitize@url \@href}%
\providecommand \@href[1]{\@@startlink{#1}\@@href}%
\providecommand \@@href[1]{\endgroup#1\@@endlink}%
\providecommand \@sanitize@url [0]{\catcode `\\12\catcode `\$12\catcode
  `\&12\catcode `\#12\catcode `\^12\catcode `\_12\catcode `\%12\relax}%
\providecommand \@@startlink[1]{}%
\providecommand \@@endlink[0]{}%
\providecommand \url  [0]{\begingroup\@sanitize@url \@url }%
\providecommand \@url [1]{\endgroup\@href {#1}{\urlprefix }}%
\providecommand \urlprefix  [0]{URL }%
\providecommand \Eprint [0]{\href }%
\providecommand \doibase [0]{http://dx.doi.org/}%
\providecommand \selectlanguage [0]{\@gobble}%
\providecommand \bibinfo  [0]{\@secondoftwo}%
\providecommand \bibfield  [0]{\@secondoftwo}%
\providecommand \translation [1]{[#1]}%
\providecommand \BibitemOpen [0]{}%
\providecommand \bibitemStop [0]{}%
\providecommand \bibitemNoStop [0]{.\EOS\space}%
\providecommand \EOS [0]{\spacefactor3000\relax}%
\providecommand \BibitemShut  [1]{\csname bibitem#1\endcsname}%
\let\auto@bib@innerbib\@empty
%</preamble>
\bibitem [{\citenamefont {Kuramoto}(1984)}]{Kuramoto-84}%
  \BibitemOpen
  \bibfield  {author} {\bibinfo {author} {\bibfnamefont {Y.}~\bibnamefont
  {Kuramoto}},\ }\href@noop {} {\emph {\bibinfo {title} {Chemical Oscillations,
  Waves and Turbulence}}}\ (\bibinfo  {publisher} {Springer},\ \bibinfo
  {address} {Berlin},\ \bibinfo {year} {1984})\BibitemShut {NoStop}%
\bibitem [{\citenamefont {Pikovsky}\ \emph {et~al.}(2003)\citenamefont
  {Pikovsky}, \citenamefont {Rosenblum},\ and\ \citenamefont
  {Kurths}}]{pikovsky2003synchronization}%
  \BibitemOpen
  \bibfield  {author} {\bibinfo {author} {\bibfnamefont {A.}~\bibnamefont
  {Pikovsky}}, \bibinfo {author} {\bibfnamefont {M.}~\bibnamefont {Rosenblum}},
  \ and\ \bibinfo {author} {\bibfnamefont {J.}~\bibnamefont {Kurths}},\
  }\href@noop {} {\emph {\bibinfo {title} {Synchronization: {A} universal
  concept in nonlinear sciences}}},\ \bibinfo {number} {12}\ (\bibinfo
  {publisher} {Cambridge University Press},\ \bibinfo {year}
  {2003})\BibitemShut {NoStop}%
\bibitem [{\citenamefont {Strogatz}(2004)}]{strogatz2004sync}%
  \BibitemOpen
  \bibfield  {author} {\bibinfo {author} {\bibfnamefont {S.}~\bibnamefont
  {Strogatz}},\ }\href@noop {} {\emph {\bibinfo {title} {Sync: The emerging
  science of spontaneous order}}}\ (\bibinfo  {publisher} {Penguin UK},\
  \bibinfo {year} {2004})\BibitemShut {NoStop}%
\bibitem [{\citenamefont {Osipov}\ \emph {et~al.}(2007)\citenamefont {Osipov},
  \citenamefont {Kurths},\ and\ \citenamefont {Zhou}}]{Osipov-Kurths-Zhou-07}%
  \BibitemOpen
  \bibfield  {author} {\bibinfo {author} {\bibfnamefont {G.}~\bibnamefont
  {Osipov}}, \bibinfo {author} {\bibfnamefont {J.}~\bibnamefont {Kurths}}, \
  and\ \bibinfo {author} {\bibfnamefont {C.}~\bibnamefont {Zhou}},\ }\href@noop
  {} {\emph {\bibinfo {title} {Synchronization in Oscillatory Networks}}}\
  (\bibinfo  {publisher} {Springer-Verlag},\ \bibinfo {address} {Berlin
  Heidelberg},\ \bibinfo {year} {2007})\BibitemShut {NoStop}%
\bibitem [{\citenamefont {Bergner}\ \emph {et~al.}(2012)\citenamefont
  {Bergner}, \citenamefont {Frasca}, \citenamefont {Sciuto}, \citenamefont
  {Buscarino}, \citenamefont {Ngamga}, \citenamefont {Fortuna},\ and\
  \citenamefont {Kurths}}]{bergner2012RS}%
  \BibitemOpen
  \bibfield  {author} {\bibinfo {author} {\bibfnamefont {A.}~\bibnamefont
  {Bergner}}, \bibinfo {author} {\bibfnamefont {M.}~\bibnamefont {Frasca}},
  \bibinfo {author} {\bibfnamefont {G.}~\bibnamefont {Sciuto}}, \bibinfo
  {author} {\bibfnamefont {A.}~\bibnamefont {Buscarino}}, \bibinfo {author}
  {\bibfnamefont {E.~J.}\ \bibnamefont {Ngamga}}, \bibinfo {author}
  {\bibfnamefont {L.}~\bibnamefont {Fortuna}}, \ and\ \bibinfo {author}
  {\bibfnamefont {J.}~\bibnamefont {Kurths}},\ }\href@noop {} {\bibfield
  {journal} {\bibinfo  {journal} {Physical Review E}\ }\textbf {\bibinfo
  {volume} {85}},\ \bibinfo {pages} {026208} (\bibinfo {year}
  {2012})}\BibitemShut {NoStop}%
\bibitem [{\citenamefont {Minati}(2015)}]{minati2015remote}%
  \BibitemOpen
  \bibfield  {author} {\bibinfo {author} {\bibfnamefont {L.}~\bibnamefont
  {Minati}},\ }\href@noop {} {\bibfield  {journal} {\bibinfo  {journal} {Chaos:
  An Interdisciplinary Journal of Nonlinear Science}\ }\textbf {\bibinfo
  {volume} {25}},\ \bibinfo {pages} {123107} (\bibinfo {year}
  {2015})}\BibitemShut {NoStop}%
\bibitem [{\citenamefont {Karakaya}\ \emph {et~al.}(2019)\citenamefont
  {Karakaya}, \citenamefont {Minati}, \citenamefont {Gambuzza},\ and\
  \citenamefont {Frasca}}]{karakaya2019fading}%
  \BibitemOpen
  \bibfield  {author} {\bibinfo {author} {\bibfnamefont {B.}~\bibnamefont
  {Karakaya}}, \bibinfo {author} {\bibfnamefont {L.}~\bibnamefont {Minati}},
  \bibinfo {author} {\bibfnamefont {L.~V.}\ \bibnamefont {Gambuzza}}, \ and\
  \bibinfo {author} {\bibfnamefont {M.}~\bibnamefont {Frasca}},\ }\href@noop {}
  {\bibfield  {journal} {\bibinfo  {journal} {Physical Review E}\ }\textbf
  {\bibinfo {volume} {99}},\ \bibinfo {pages} {052301} (\bibinfo {year}
  {2019})}\BibitemShut {NoStop}%
\bibitem [{\citenamefont {Vuksanovi{\'c}}\ and\ \citenamefont
  {H{\"o}vel}(2014)}]{vuksanovic2014functional}%
  \BibitemOpen
  \bibfield  {author} {\bibinfo {author} {\bibfnamefont {V.}~\bibnamefont
  {Vuksanovi{\'c}}}\ and\ \bibinfo {author} {\bibfnamefont {P.}~\bibnamefont
  {H{\"o}vel}},\ }\href@noop {} {\bibfield  {journal} {\bibinfo  {journal}
  {NeuroImage}\ }\textbf {\bibinfo {volume} {97}},\ \bibinfo {pages} {1}
  (\bibinfo {year} {2014})}\BibitemShut {NoStop}%
\bibitem [{\citenamefont {Vlasov}\ and\ \citenamefont
  {Bifone}(2017)}]{vlasov2017hub}%
  \BibitemOpen
  \bibfield  {author} {\bibinfo {author} {\bibfnamefont {V.}~\bibnamefont
  {Vlasov}}\ and\ \bibinfo {author} {\bibfnamefont {A.}~\bibnamefont
  {Bifone}},\ }\href@noop {} {\bibfield  {journal} {\bibinfo  {journal}
  {Scientific reports}\ }\textbf {\bibinfo {volume} {7}},\ \bibinfo {pages} {1}
  (\bibinfo {year} {2017})}\BibitemShut {NoStop}%
\bibitem [{\citenamefont {Gambuzza}\ \emph {et~al.}(2013)\citenamefont
  {Gambuzza}, \citenamefont {Cardillo}, \citenamefont {Fiasconaro},
  \citenamefont {Fortuna}, \citenamefont {G{\'o}mez-Gardenes},\ and\
  \citenamefont {Frasca}}]{gambuzza2013analysis}%
  \BibitemOpen
  \bibfield  {author} {\bibinfo {author} {\bibfnamefont {L.~V.}\ \bibnamefont
  {Gambuzza}}, \bibinfo {author} {\bibfnamefont {A.}~\bibnamefont {Cardillo}},
  \bibinfo {author} {\bibfnamefont {A.}~\bibnamefont {Fiasconaro}}, \bibinfo
  {author} {\bibfnamefont {L.}~\bibnamefont {Fortuna}}, \bibinfo {author}
  {\bibfnamefont {J.}~\bibnamefont {G{\'o}mez-Gardenes}}, \ and\ \bibinfo
  {author} {\bibfnamefont {M.}~\bibnamefont {Frasca}},\ }\href@noop {}
  {\bibfield  {journal} {\bibinfo  {journal} {Chaos: An Interdisciplinary
  Journal of Nonlinear Science}\ }\textbf {\bibinfo {volume} {23}},\ \bibinfo
  {pages} {043103} (\bibinfo {year} {2013})}\BibitemShut {NoStop}%
\bibitem [{\citenamefont {Sakaguchi}\ and\ \citenamefont
  {Kuramoto}(1986)}]{sakaguchi1986soluble}%
  \BibitemOpen
  \bibfield  {author} {\bibinfo {author} {\bibfnamefont {H.}~\bibnamefont
  {Sakaguchi}}\ and\ \bibinfo {author} {\bibfnamefont {Y.}~\bibnamefont
  {Kuramoto}},\ }\href@noop {} {\bibfield  {journal} {\bibinfo  {journal}
  {Progress of Theoretical Physics}\ }\textbf {\bibinfo {volume} {76}},\
  \bibinfo {pages} {576} (\bibinfo {year} {1986})}\BibitemShut {NoStop}%
\bibitem [{\citenamefont {Gengel}\ \emph {et~al.}(2020)\citenamefont {Gengel},
  \citenamefont {Teichmann}, \citenamefont {Rosenblum},\ and\ \citenamefont
  {Pikovsky}}]{gengelphasereduction}%
  \BibitemOpen
  \bibfield  {author} {\bibinfo {author} {\bibfnamefont {E.}~\bibnamefont
  {Gengel}}, \bibinfo {author} {\bibfnamefont {E.}~\bibnamefont {Teichmann}},
  \bibinfo {author} {\bibfnamefont {M.}~\bibnamefont {Rosenblum}}, \ and\
  \bibinfo {author} {\bibfnamefont {A.}~\bibnamefont {Pikovsky}},\ }\href@noop
  {} {\bibfield  {journal} {\bibinfo  {journal} {Journal of Physics:
  Complexity}\ }\textbf {\bibinfo {volume} {2}},\ \bibinfo {pages} {015005}
  (\bibinfo {year} {2020})}\BibitemShut {NoStop}%
\bibitem [{\citenamefont {Qin}\ \emph {et~al.}(2020)\citenamefont {Qin},
  \citenamefont {Cao}, \citenamefont {Anderson}, \citenamefont {Bassett},\ and\
  \citenamefont {Pasqualetti}}]{qin2020mediated}%
  \BibitemOpen
  \bibfield  {author} {\bibinfo {author} {\bibfnamefont {Y.}~\bibnamefont
  {Qin}}, \bibinfo {author} {\bibfnamefont {M.}~\bibnamefont {Cao}}, \bibinfo
  {author} {\bibfnamefont {B.~D.}\ \bibnamefont {Anderson}}, \bibinfo {author}
  {\bibfnamefont {D.~S.}\ \bibnamefont {Bassett}}, \ and\ \bibinfo {author}
  {\bibfnamefont {F.}~\bibnamefont {Pasqualetti}},\ }\href@noop {} {\bibfield
  {journal} {\bibinfo  {journal} {IEEE Control Systems Letters}\ }\textbf
  {\bibinfo {volume} {5}},\ \bibinfo {pages} {767} (\bibinfo {year}
  {2020})}\BibitemShut {NoStop}%
\bibitem [{\citenamefont {Qin}\ \emph {et~al.}(2018)\citenamefont {Qin},
  \citenamefont {Kawano},\ and\ \citenamefont {Cao}}]{qin2018stability}%
  \BibitemOpen
  \bibfield  {author} {\bibinfo {author} {\bibfnamefont {Y.}~\bibnamefont
  {Qin}}, \bibinfo {author} {\bibfnamefont {Y.}~\bibnamefont {Kawano}}, \ and\
  \bibinfo {author} {\bibfnamefont {M.}~\bibnamefont {Cao}},\ }in\ \href@noop
  {} {\emph {\bibinfo {booktitle} {2018 IEEE Conference on Decision and Control
  (CDC)}}}\ (\bibinfo {organization} {IEEE},\ \bibinfo {year} {2018})\ pp.\
  \bibinfo {pages} {5209--5214}\BibitemShut {NoStop}%
\bibitem [{\citenamefont {Nicosia}\ \emph {et~al.}(2013)\citenamefont
  {Nicosia}, \citenamefont {Valencia}, \citenamefont {Chavez}, \citenamefont
  {D{\'\i}az-Guilera},\ and\ \citenamefont {Latora}}]{nicosia2013remote}%
  \BibitemOpen
  \bibfield  {author} {\bibinfo {author} {\bibfnamefont {V.}~\bibnamefont
  {Nicosia}}, \bibinfo {author} {\bibfnamefont {M.}~\bibnamefont {Valencia}},
  \bibinfo {author} {\bibfnamefont {M.}~\bibnamefont {Chavez}}, \bibinfo
  {author} {\bibfnamefont {A.}~\bibnamefont {D{\'\i}az-Guilera}}, \ and\
  \bibinfo {author} {\bibfnamefont {V.}~\bibnamefont {Latora}},\ }\href@noop {}
  {\bibfield  {journal} {\bibinfo  {journal} {Physical Review Letters}\
  }\textbf {\bibinfo {volume} {110}},\ \bibinfo {pages} {174102} (\bibinfo
  {year} {2013})}\BibitemShut {NoStop}%
\bibitem [{Note1()}]{Note1}%
  \BibitemOpen
  \bibinfo {note} {Notice that our Eq.~(\ref {eqn:SLequations}) represents a
  particular case of a more general setup studied in~\cite
  {gengelphasereduction}}\BibitemShut {NoStop}%
\bibitem [{Note2()}]{Note2}%
  \BibitemOpen
  \bibinfo {note} {In the following, we always consider frequencies of the
  peripheral oscillators to be close while the frequency of the hub is
  essentially different.}\BibitemShut {Stop}%
\bibitem [{Note3()}]{Note3}%
  \BibitemOpen
  \bibinfo {note} {We sweep the parameter space to determine the state of the
  system using efficient techniques. We find RS in the SL equations~\protect
  \textup {\hbox {\mathsurround \z@ \protect \normalfont (\ignorespaces \ref
  {eqn:SLequations}\unskip \@@italiccorr )}} by looking for a limit cycle
  solution where $\gamma _{12}$ is unbounded while $\gamma _{13}$ is bounded
  using a shooting method. To detect RS using the phase reduction equations
  exactly, we numerically construct the Poincar\'e map described in Sec. \ref
  {sec:RS_theoretical} by simulating Eq.~\protect \textup {\hbox {\mathsurround
  \z@ \protect \normalfont (\ignorespaces \ref
  {eqn:first_order_gamma13/gamma12}\unskip \@@italiccorr )}} (or its
  second-order counterpart) and check the presence of a fixed point. Regions of
  CS were computed using direct numerical simulations; we mark the points in
  the parameter space that resulted in $\Omega _1=\Omega _2=\Omega _3$ up to a
  tolerance of $10^{-4}$.}\BibitemShut {Stop}%
\bibitem [{Note4()}]{Note4}%
  \BibitemOpen
  \bibinfo {note} {The borderline of the RS transition is given by the
  condition when inequalities (\ref {eqn:first_order_RS_condition},\ref
  {eqn:second_order_RS_condition}) turn to equalities.}\BibitemShut {Stop}%
\bibitem [{\citenamefont {Kralemann}\ \emph {et~al.}(2011)\citenamefont
  {Kralemann}, \citenamefont {Pikovsky},\ and\ \citenamefont
  {Rosenblum}}]{Kralemann-Pikovsky-Rosenblum-11}%
  \BibitemOpen
  \bibfield  {author} {\bibinfo {author} {\bibfnamefont {B.}~\bibnamefont
  {Kralemann}}, \bibinfo {author} {\bibfnamefont {A.}~\bibnamefont {Pikovsky}},
  \ and\ \bibinfo {author} {\bibfnamefont {M.}~\bibnamefont {Rosenblum}},\
  }\href@noop {} {\bibfield  {journal} {\bibinfo  {journal} {Chaos}\ }\textbf
  {\bibinfo {volume} {21}},\ \bibinfo {pages} {025104} (\bibinfo {year}
  {2011})}\BibitemShut {NoStop}%
\bibitem [{\citenamefont {Lacerda}\ \emph {et~al.}(2019)\citenamefont
  {Lacerda}, \citenamefont {Freitas},\ and\ \citenamefont
  {Macau}}]{lacerda2019multistable}%
  \BibitemOpen
  \bibfield  {author} {\bibinfo {author} {\bibfnamefont {J.}~\bibnamefont
  {Lacerda}}, \bibinfo {author} {\bibfnamefont {C.}~\bibnamefont {Freitas}}, \
  and\ \bibinfo {author} {\bibfnamefont {E.}~\bibnamefont {Macau}},\
  }\href@noop {} {\bibfield  {journal} {\bibinfo  {journal} {Applied
  Mathematical Modelling}\ }\textbf {\bibinfo {volume} {69}},\ \bibinfo {pages}
  {453} (\bibinfo {year} {2019})}\BibitemShut {NoStop}%
\bibitem [{\citenamefont {Miller}(2006)}]{miller2006applied}%
  \BibitemOpen
  \bibfield  {author} {\bibinfo {author} {\bibfnamefont {P.~D.}\ \bibnamefont
  {Miller}},\ }\href@noop {} {\emph {\bibinfo {title} {Applied asymptotic
  analysis}}},\ Vol.~\bibinfo {volume} {75}\ (\bibinfo  {publisher} {American
  Mathematical Soc.},\ \bibinfo {year} {2006})\BibitemShut {NoStop}%
\end{thebibliography}

\end{document}